\documentclass[doublecol,graphicx]{epl2} 
\usepackage{graphicx}
\usepackage{amsmath}
\usepackage{amsfonts}
\usepackage{amssymb}
\usepackage{amsbsy}
\usepackage{mathptmx}

\newcommand{\ag}{Ann\ Geo\-phy\-si\-c\ae}

\newcommand{\jgr}{J\ Geo\-phys Res}

\newcommand{\prl}{Phys\ Rev\ Lett}

\newcommand{\pss}{Planet\ Space Sci}

\newcommand{\pfl}{Phys Fluids}

\newcommand{\pop}{Phys\ Plasmas}

\def\|{{\sss\parallel}}

\def\sss{\scriptscriptstyle}
\title{On the {linear mirror mode threshold in} high-$\beta$ space plasma}
\shorttitle{Mirror mode thresholds} 

\author{R. A. Treumann\inst{1,2} \thanks{Visiting the International Space Science Institute, Bern, Switzerland} \and O. D. Constantinescu$^3$}
\shortauthor{R. A. Treumann, \and O. D. Constantinescu}

\institute{ 
  \inst{1} Department of Geophysics and Environmental Sciences, Munich University, Munich, Germany\\                   
  \inst{2} Department of Physics and Astronomy, Dartmouth College, Hanover, NH 03755\\
  \inst{3} Institute of Geophysics and extraterrestrial Physics, Technical University, Braunschweig, Germany
  }
 

\abstract{The linear threshold condition of the mirror mode including weak drift and finite Larmor radius effects  is analysed and shown that it can be given to arbitrary precision in closed analytical form. Combined with observations of mirror modes in the magnetosheath it is shown that only magnetic `dips' are produced by the mirror instability. A semi-experimental threshold condition is derived. The occasionally observed so-called `peaks' cannot be attributed to mirror modes but result probably from nonlinear evolution of the compressive fast mode.}

\begin{document}

\maketitle
\section{Introduction}
The magnetic mirror mode is one of the most interesting plasma modes in a collisionless anisotropic-pressure high-$\beta$ plasma. It is practically non-oscillating and occurs  when the perpendicular (thermal) plasma (ion) pressure $P_\perp >P_\|$ sufficiently exceeds the parallel pressure. This mode has originally been inferred from simple fluid theory \cite{chandra1958} and in the following decades has been investigated more closely in ever higher precision (cf., e.g., \cite{hasegawa1969,hasegawa1975,pokh2002}). It is of particular interest because being of nearly zero frequency and being capable of assuming large amplitudes it may serve as energy source in the turbulent cascade that leads to low frequency magnetic plasma turbulence \cite{biskamp2003}. 

Space plasma observations in various regions, planetary magnetosheaths, the inner terrestrial magnetosphere, near comets, and even in the solar wind provided evidence for the presence of mirror modes under the conditions prevalent in these space plasmas. (While a comprehensive critical review of the observations and theory is in high demand, it is not the place for it here. The reader is {therefore} referred to some less contemporary reviews, as e.g., \cite{schwartz1996}.) Still, even though the mirror mode seems to be a very simple basic plasma eigenmode, controversy remains about its mechanism and physical realisation. Several competing linear and nonlinear theories and sometimes controversial opinions have been put forward. Numerical simulations have also been performed but are not in the position to clarify the case. It is, however, clear that for the mirror mode to occur as an eigenmode of anisotropic pressure plasmas it should evolve from a linear plasma instability starting from the prescribed anisotropic initial conditions. These can be very different, either assuming a homogeneous plasma, a streaming plasma, a weakly or strongly inhomogeneous plasma, a two-component plasma consisting of protons and electrons or a plasma with an admixture of heavy ions, and depending on the undisturbed ion distribution function and possibly even electron temperature and electron pressure anisotropy. In addition, since the mode is a long wavelength mode being of macroscopic size both along and perpendicular to the external magnetic field, one might need to include the spatial boundaries of the region where the modes have been observed. This applies in particular to the Earth's magnetosheath which is bounded by the bow shock and the magnetopause. To account for all these initial conditions makes little sense. Hence one choses one or the other sufficiently simple initial condition. In order to clarify the basic state of the linear instability it is most instructive to investigate the weakly inhomogeneous unbounded case in a bi-Maxwellian two-component plasma. 
The present Letter intends to clarify what can be learned from the available most elaborate {\it linear} mirror mode theory. We also have in view its application to regions in space where mirror modes have been observed but definitely restrict to the linear treatment only. Some conclusions can be drawn from its comparison with recent observations in the Earth's magnetosheath {\cite{soucek2008}}. 

{\section{Instability threshold condition} The most complete kinetic derivation of the linear growth rate of the mirror mode including a weak  inhomogeneity ${\bf B}(y)=B_0(1-\alpha y) \hat z$ in an anisotropic high-$\beta$ plasma under the condition that the plasma is bi-Maxwellian (at this stage of the theory an unimportant restriction) has been given by Pokhotelov et al. {\cite{pokh1976,pokh1985,pokh2002,pokh2004,pokh2005}}.  For the inhomogeneity to be weak only, the inhomogeneity parameter is restricted as $|\alpha|\ll k_\perp$, where ${\bf k}=(k_\perp,k_\|)$ is the wave number of a plane wave. Since the inhomogeneity is strictly perpendicular to the magnetic field no such condition holds for $k_\|$. This contrasts other claims  \cite{hellinger2008} of an apparent impossibility to derive an instability threshold. In fact, $k_\|$ can become arbitrarily small. The smallness condition on the inhomogeneity also comes up for the decoupling of the mirror mode from the kinetic Alfv\'en wave which on scales of the order of the ion inertial length $\lambda_i=c/\omega_{pi}$ is another (non-zero frequency) eigenmode in high-$\beta$ plasma. This coupling takes place at steep plasma gradients, i.e. it should play a role in the nonlinear evolution of the mirror mode whenever it grows to assume large amplitudes and evolves steep gradients. Since it has been shown that maximum growth of the mirror mode occurs at $k_\perp\rho_i={\rm O}(1)$, with $\rho_i=v_{i\perp}/\omega_{ci}$ the ion gyroradius \cite{pokh1985,pokh2004}  ($v_{i\perp}, v_{i\|}$ are the respective perpendicular and parallel thermal velocities of the ions, and $\omega_{ci}$ is the ion cyclotron frequency), this condition also implies that $|\alpha\rho_i|\ll 1$.  Linear theory then yields for the real frequency of the unstably excited mode
\begin{equation}\label{eqomega}
\frac{\omega({\bf k})}{k_\perp v_{i\perp}}=\frac{3}{2\beta_\perp}\left(A\beta_\perp -\frac{{2}}{3}\right)\frac{\alpha\rho_i}{A+1}
\end{equation}
Here we introduced the pressure (or temperature) anisotropy $A=(T_{\perp}/T_{\|})-1>0$ which, for the mirror mode is always assumed to be positive. {(Since for the mirror mode  $\beta_\perp$ is important, with $\beta_\|$ being dynamically irrelevant, all expressions in this Letter are given in terms of $A$ and $\beta_\perp$.)} Under these assumptions the complete expression for the linear growth rate becomes
\begin{equation}\label{eqgamma}
\frac{\gamma}{\gamma_0}=A\beta_\perp -1-\frac{3}{2}(k_\perp\rho_i)^2-\frac{k_\|^2}{k_\perp^2}\left(1+\frac{\beta_\|}{2}A\right)
\end{equation}
$\gamma_0=(2/\pi)^\frac{1}{2}\left\{|k_\||v_{i\|}/\beta_\perp(A+1)\right\}$ is a positive factor. Clearly $\gamma\to 0$ for $k_\|\to 0$. This expression is independent of $\alpha$. In the linear approximation the inhomogeneity enters only in the real frequency. 

Instability arises for positive right hand side in Eq. (\ref{eqgamma}). The first two terms, put to zero, yield the ordinary mirror threshold condition $A\beta_\perp\geq 1$ for homogeneous plasma \cite{hasegawa1969}, neglecting the finite Larmor radius effect in the third term. The fourth term gives a further correction obtained in the above reference. Both, the finite Larmor radius contribution and this term reduce the growth rate and increase the threshold. Clearly the classical condition is an oversimplification of the real conditions even when neglecting the finite Larmor radius effect. Setting the right hand side in $\gamma$ to zero one obtains the threshold condition for marginal instability of the mirror mode. 

Before drawing any conclusion about stability it is of interest to investigate the real frequency. This is clearly different from zero only in the case of inhomogeneity. Thus any modes have finite phase and group velocities in the plasma frame only when the inhomogeneity is taken into account or, otherwise, when the plasma is moving. However, Eq. (\ref{eqomega}) suggests that there are two possible modes whether $A\beta_\perp\gtrless\frac{{2}}{3}$ such that it seems to be necessary in the growth rate to consider these two cases separately. {However, this is not the case as linearly growing waves in the plasma frame can be obtained only when the larger sign holds, as will become clear below. Growing drift-mirror modes propagate down the gradient in the direction of decreasing magnetic field, which intuitively is reasonable because of the decreasing magnetic pressure and stresses that support wave growth.}

Defining $X=A+1$, it is easy to show that for the positive $A >\frac{{2}}{3} $ the marginal stability threshold condition becomes
\begin{equation}\label{eqX}
X^2-X\left[1+\frac{1}{\beta_\perp}\left(1+\frac{k_\|^2}{k_\perp^2}+\frac{3}{2}k_\perp^2\rho_i^2\right)\right]-\frac{1}{3\beta_\perp}\frac{k_\|^2}{k_\perp^2}>0
\end{equation}
where we replaced $\beta_\| =  \beta_\perp /(A+1)$. The expression in the large bracket is positive definite. Hence, there is just one solution with $X>0$ which is given by
\begin{equation}
X>\frac{1}{2}[\cdots]\left\{1+\left(1+\frac{{4}}{3\beta_\perp}\frac{k_\|^2}{[\cdots]^2k_\perp^2}\right)^\frac{1}{2}\right\}
\end{equation}
Here $[\cdots]$ stands for the large bracket in Eq. (\ref{eqX}). One immediately realises that a slightly finite Larmor radius modified mirror threshold is obtained for small $k_\|^2\ll k_\perp^2$ which reads
\begin{equation}
A\beta_\perp>1+\frac{3}{2}k_\perp^2\rho_i^2 + \dots, \qquad k_\|^2\ll k_\perp^2
\end{equation}
with positive definite correction term
\begin{equation}
\dots =\frac{1}{3}\frac{k_\|^2}{k_\perp^2}\left[1+\frac{1}{\beta_\perp}\left(1+\frac{k_\|^2}{k_\perp^2}+\frac{3}{2}k_\perp^2\rho_i^2\right) \right]^{-1}
\end{equation}
{The larger $\beta_\perp$ the larger is the effect of the term inside the bracket on the threshold, and for very large $\beta_\perp\gg 1$ the correction term reduces to $k_\|^2/3k_\perp^2$. This case is unambiguously identified as the linear drift mirror mode. (The condition $k_\|^2\ll k_\perp^2$ is natural for linear mirror modes for which magnetic tension inhibits oblateness. This holds even in the nonlinear case.)  Both, the finite Larmor radius effect and the parallel wave number term act stabilising.} Further improvements can be done by including finite electron temperature effects or anisotropic non-Maxwellian distribution functions, but do not change the conclusion.

Turning to the case with $A<\frac{{2}}{3}$ which describes {uphill moving} waves, the threshold condition for instability becomes
\begin{equation}\label{eqneg}
{1+\frac{9}{2}k_\perp^2\rho_i^2+\frac{3k_\|^2}{k_\perp^2}\left(1+\frac{\beta_\perp}{2}\frac{A}{A+1}\right)<0}
\end{equation}
{This can be satisfied only when $A<0$, a case that is not covered by the derivation of the mirror mode but applies to Alfv\'en firehose waves. Hence, in this range mirror waves do not grow. Uphill moving mirror modes are linearly stable. This does not preclude that they may not become parametrically excited whenever the parametric growth rate exceeds the linear damping rate in Eqs. (\ref{eqgamma}) and (\ref{eqneg}). This may happen when mirror waves evolve nonlinearly, generate sidebands and cascade into turbulence. 

Thus the sole case of linearly growing mirror modes is covered by downhill propagating waves which, as is well known from linear analysis \cite{pokh1976,pokh1985,pokh2002} take their free energy from resonance with the bulk of low parallel energy ions of velocity $v_\|\approx 0$. We note in passing that this bulk resonance makes the difference between mirror and electromagnetic cyclotron modes which otherwise grow under the same condition of positive $A>0$. The latter waves are in resonance with the high energy particle component of resonant energy ${\cal E}_{res}\gtrsim m_i(\omega_{ci}-\omega)^2/2k_\|^2={\cal E}_B(1-\omega/\omega_{ci})^2/k_\|^2\lambda_i^2$, where $\omega=\omega(\bf k)$ is the electromagnetic ion cyclotron dispersion relation, $\lambda_i=c/\omega_{pi}$ the ion  inertial length, and ${\cal E}_B=m_iV_A^2/2$ the Alfv\'en energy. Thus the two instabilities are about independent. They may or may not coexist. (Under low-$\beta$ conditions the ranges of mirror and electromagnetic ion cyclotron instabilities have recently been investigated  \cite{treu2008}).}

\section{Application to magnetosheath observations}
In linear theory there is no difference between the mirror and drift-mirror instabilities in their growth rates. If streaming with velocity ${\bf{V}}$ would be included then the mirror mode would of course as well become oscillating with Doppler frequency $\omega={\bf k\cdot V}$ \cite{sahr2004} when transforming from the plasma to the laboratory frame. 

Observation of the mirror mode in its linear state is hardly possible. Wave spectra measured in the magnetosheath \cite{sahr2004} have been interpreted as indicating the mirror dispersion relation. However, the errors are very large and the linear approximation is questionable as the observed waves have amplitudes $b/B_0$ reaching from several \% to several 10\% indicating that one is dealing with nonlinear structures which develop steep spatial gradients in the magnetic field and plasma density and $\beta$, thereby violating the assumptions of linear theory and coupling the mirror mode to the kinetic Alfv\'en wave.  Observations in the magnetosheath in addition indicated (e.g. \cite{hill1995}) that mirror structures seem to undergo an interesting spatial transformation between the bow shock, the outer boundary of the magnetosheath, and the magnetopause, the inner boundary of the magnetosheath. In the outer part of the magnetosheath the mode seems to nonlinearly develop into a sequence of `magnetic walls' (peaks), while closer to the magnetopause it occurs as `magnetic holes' (dips). Admittedly, it is problematic to decide experimentally between these two forms simply because the definition of the background magnetic field in a highly structured (or spatially fluctuating) magnetic field is difficult to determine. 

\subsection{{Experimentally determined threshold}} A recently performed more sophisticated and also more precise analysis \cite{soucek2008} taking advantage of  spatial Cluster four-point measurements in the magnetosheath confirms these observations. Soucek et al. \cite{soucek2008} construct a quantity they call `peakness' ${\cal P}$ with the help of which they can approximately distinguish between `peaks' ${\cal P}>0$ and `dips' ${\cal P}<0$. This quantity correlates with $\beta_\|$ and $\beta_\perp A$ (their Figure 2). If we eliminate ${\cal P}$ from their data (which they provide as scatter plots) it is possible to construct the statistical relation $\beta_\| \approx 0.6\,\exp\,(0.3\,\beta_\perp A)$. Expressing $\beta_\|$ through the anisotropy this yields a relation {between} $A$ and $\beta_\perp$ which is of the form
\begin{equation}
\beta_\perp A\approx 0.5 +\ln\left(\frac{\beta_\perp}{A+1}\right)
\end{equation}
This expression is derived without distinguishing between `peaks' and `dips'. It should be refined by considering these two {data} sets separately which might lead (in addition to a precision of the numerical coefficients) to two separate relations {of which only the one for `dips' will apply to the mirror mode. This will become clear below.}

In spite of {these reservations}, the above relation is interesting as it {resembles} a `nonlinear threshold condition' for the mirror mode {in close similarity to the simple linear threshold condition}. The logarithmic nonlinearity in this expression is a consequence of the notoric observation of mirror modes in their nonlinear state. However, {the nonlinearity is logarithmic} and is thus weak. The main difference is in the statistically uncertain numerical term 0.5 on the right. {(Unfortunately, in \cite{soucek2008} no lower limit on $|{\cal P}|$ has been set such that the correlations are infected by the accumulation of irrelevant cases around ${\cal P}\approx 0$.)} 

\begin{figure}[t!]
\centerline{\includegraphics[width=0.4\textwidth,clip=]{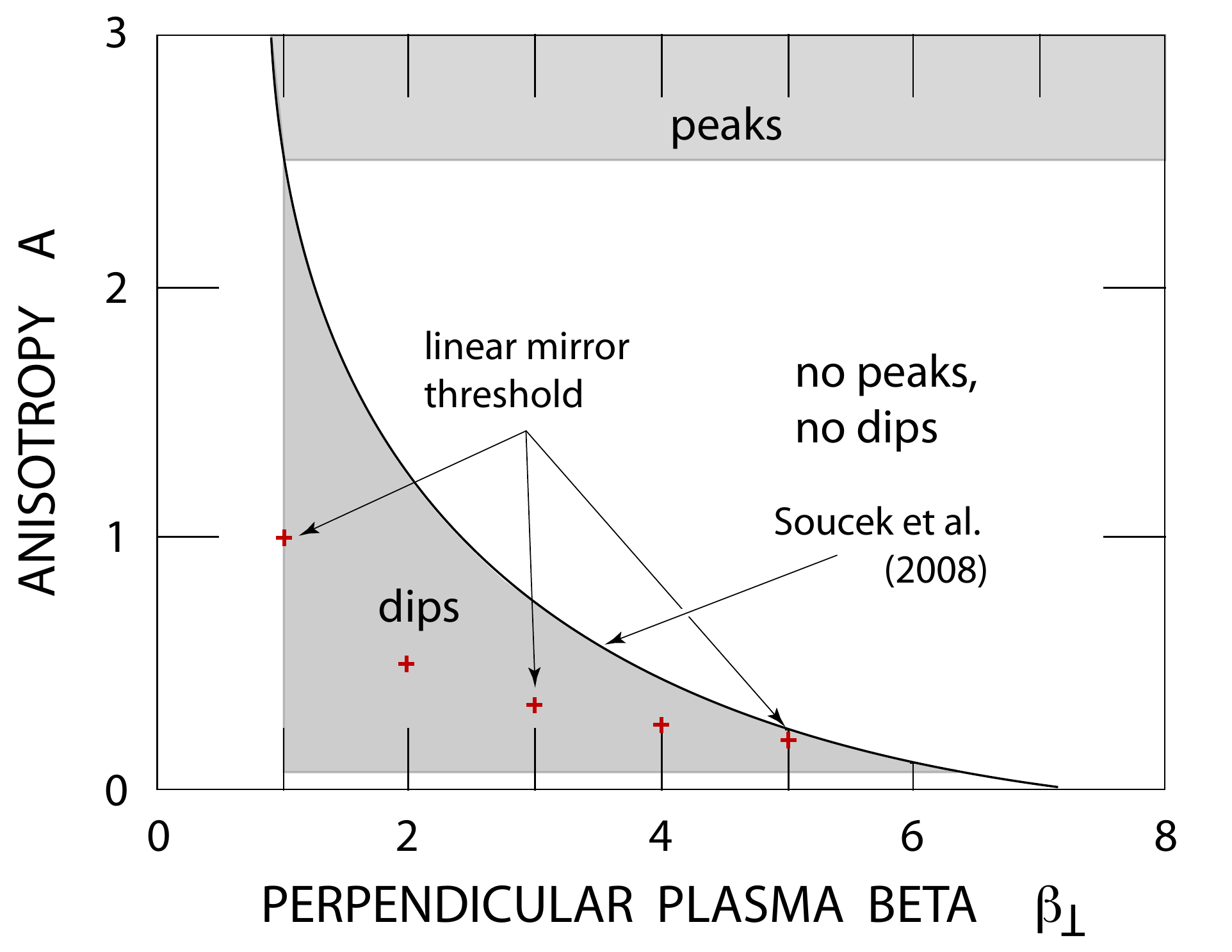} }
\caption[ ]
{\footnotesize The ranges of mirror `dips' and `peaks' in the plane of the pressure anisotropy $A$ and $\beta_\perp$ for the Earth's magnetosheath based on the observations of \cite{soucek2008} and purely linear theory. This representation makes use of the experimentally determined approximate boundary between `dips' and `peaks'  (black line) {and the theoretical mirror threshold condition. For comparison the crosses indicate the linear threshold condition for the mirror mode. It is obvious that the mirror mode cannot produce any of the observed `peaks' above the experimental curve for anisotropies below $A\sim 2.5$. On the other hand, the experimental threshold allows for `dips' both, above and below the linear mirror threshold}.}\label{peaksdips}
\vspace{-0.3cm}\end{figure}

\subsection{{`Peaks' or `dips'? Only `dips' survive}} Analysing their distribution of `dips' and `peaks', the authors of \cite{soucek2008} found an approximate relation $T_\perp/T_\|\approx2.2/\beta_\|^{\,0.4}$ between the temperature anisotropy and $\beta_\|$ for the boundary between the regions of `dips' and `peaks' \cite{soucek2008}. For values below this approximate boundary curve they predominantly observe `dips', while above it they predominantly identify `peaks'. Rewritten in our terms for the `dips' this relation becomes
\begin{equation}\label{condpos}
A<(3.5/\beta_\perp^{\,0.64})-1 
\end{equation}
The numbers in this expression (in particular the power on $\beta_\perp$) are quite uncertain, however, due to the large scatter in the data \cite{soucek2008}. {(Note that the power of $\beta_\perp$ is very close to $\frac{2}{3}$.)} Still, this can be compared with the marginal mirror instability threshold, yielding for `dips'
\begin{equation}
1\lesssim A\beta_\perp\lesssim3.5\,\beta_\perp^{\,0.36}-\beta_\perp\sim 3.5\,\beta_\perp^\frac{1}{3}-\beta_\perp
\end{equation}
This condition can be satisfied as long as $1<\beta_\perp<6.3$, which yields for the anisotropy that $0.15<A<2.5$ for the occurrence of `dips'. Thus  both, anisotropy $A$ and  $\beta_\perp$, are restricted to small values. 

On the other hand, accounting for  `peaks', the above inequality {Eq. (\ref{condpos})} is inverted, and all anisotropies $A>2.5$ are permitted for causing `peaks'. These regions are shown graphically in Figure \ref{peaksdips} where we also have drawn the linear mirror threshold boundary (crosses).

Given that the experimentally determined condition \cite{soucek2008} holds, we apparently find that it is in accord with the approximate threshold relation of the mirror mode criterion for moderately large $\beta_\perp$ to cause `dips', {while the peaks observed by Soucek et al. \cite{soucek2008} to accumulate in the large white region above the boundary in Figure \ref{peaksdips} obviously are no mirror mode signatures.} Only peaks at large anisotropies could possibly be caused by mirror modes. However, from linear analysis this can hardly be definitively concluded to be the case.  {It is also interesting to see that the allowed region of `dip' occurrence centres around the marginal threshold condition of the linear mirror mode. It is suggestive that `dips' are found both above threshold (in the linearly unstable domain) and below threshold (in the linearly stable domain), however with both regions being quite narrow in parameter space. }

\section{Discussion} In order to reach the observed large `dip' amplitudes the mirror instability needs to enter the nonlinear state, which might have proceeded along the lines investigated for evolution near marginal stability \cite{kuz2007,pokh2008}. {The main conclusion we can draw from linear theory is that it is improbable that the mirror instability near threshold would be responsible for generating `peaks'. This conjecture could experimentally be proven by eliminating all statistically uncertain small amplitude $|{\cal P|}$ values from the data of \cite{soucek2008} and investigating the data sets for `peaks' and `dips' separately. 

On the other hand, nonlinear mirror mode theory \cite{pokh2008} and simulations \cite{calif2008,shoji2008} give a clue on the evolution of the mirror instability. It has been proposed recently in \cite{kuz2007} that near marginal stability mirror mode collapse could take place. If this happened it would completely determine the further evolution of the instability, in particular when it starts from a large amplitude disturbance. Though this theory is very attractive and beautiful, a more sophisticated  and elaborated nonlinear investigation \cite{pokh2008} has shown that the terms in the nonlinear evolution of the mirror instability that lead to collapse do in fact cancel such that collapse does not take place. Instead, the mirror mode evolves solely into `dips', in perfect agreement with the results of this investigation which are based exclusively on linear theory combined with observation \cite{soucek2008}. Saturation of the mirror mode in its nonlinear state is then provided by trapping of the resonant low-parallel energy ion component \cite{pokh2008} that causes the magnetic field to flatten out in the centre of the mirror `dips', similar to what had been found in the very high-time resolution magnetic observations of mirror modes \cite{baum1999} provided by the Equator-S spacecraft.    

What concerns the `peaks', it seems that from a theoretical point of view the mirror instability is not responsible for their evolution, neither when it evolves from thermal fluctuation level, nor when it starts from large amplitude disturbances. In the latter case the mirror mode becomes the drift mirror mode and preferentially propagates into the direction of decreasing magnetic field thereby further depressing the field. Uphill moving modes could possibly be generated by parametric instability requiring the existence of large amplitude `dips'. In this case they would be short wavelength structures such as have been found in the Equator-S recordings inside mirror bubbles. Compressions of the field should be expected to occur in contrast when a compressive wave mode is excited in the plasma. This is naturally the case when the fast mode is unstable, a case that was suggested also in \cite{pokh2008}. The observation that compressive `peaks' in the magnetosheath are found close to the bow shock hints on a connection to the shock as, in particular, at short distances from the shock the thermal anisotropy of the magnetosheath plasma has not yet built up sufficiently and the plasma is probably mirror stable as was suggested in \cite{hill1995}. From quasi-parallel shock theory one may indeed expect  that behind the quasi-parallel bow shock the shock radiates compressive waves which in the magnetosheath appear as large amplitude compressions or solitary structures on the fast mode.}}
  
{\small \acknowledgements
This research is part of a Visiting Scientist Programme at ISSI, Bern. Hospitality of ISSI  is thankfully recognised. Previous discussions and cooperations with O. Pokhotelov are highly appreciated.

}


\vspace{-0.3cm}
\parskip=0pt


\begin{thebibliography}{0}

\bibitem{chandra1958}
\Name{S A Chandrasekhar, A N Kaufman \and  K M Watson}
\REVIEW{Proc Roy Soc London Ser A}{245}{1958}{435}.

\bibitem{hasegawa1969}
\Name{A Hasegawa}
\REVIEW{\pfl}{12}{1969}{2642, doi: 10.1063/1.1692407}.

\bibitem{hasegawa1975} 
\Name{A Hasegawa} \Book{Plasma Instabilities and Nonlinear Effects} \Publ{Sprin\-ger Verlag, Berlin \Year{1975}}.

\bibitem{pokh2002} 
\Name{O A Pokhotelov et al} \REVIEW{\jgr}{107}{2002}{A092 13, doi: 10.1029/2004JA010568}.

\bibitem{biskamp2003} 
\Name{D Biskamp} \Book{Magnetohydrodynamic Turbulence} \Publ{Cambridge Un\-iversity Press, Cambridge \Year{1975}}.

\bibitem{schwartz1996}
\Name{S J Schwartz,  D Burgess \and J J Moses}
\REVIEW{\ag}{14}{1996}{1134}.

\bibitem{soucek2008}
\Name{J Soucek, E A Lucek \and I Dandouras}
\REVIEW{\jgr}{113}{2008}{A04203, doi: 10.1029/2007JA012649}.


{\bibitem{pokh1985} 
\Name{O A Pokhotelov,  V A Pilipenko \and E Amata} \REVIEW{\pss}{33}{1985}{1229, doi: 10.1016/0032-0633(85)90001-7}.
\bibitem{pokh1976} 

\Name{O A Pokhotelov \and V A Pilipenko} \REVIEW{Geomagn Aeron}{16}{1976}{296}.}

\bibitem{pokh2004} 
\Name{O A Pokhotelov, R Z Sagdeev,  M A Balikhin \and  R A Treumann} \REVIEW{\jgr}{109}{2004}{9213, doi: 10.1029/ 2004JA010568}.

\bibitem{pokh2005} 
\Name{O A Pokhotelov, M A Balikhin, R Z Sagdeev \and R A Treumann} \REVIEW{\jgr}{110}{2005}{9213, doi: 10.1029/ 2004JA010933}.

\bibitem{hellinger2008}
\Name{ P Hellinger} \REVIEW{\pop}{15}{2008}{054502, doi: 10.1063/1. 2912961}.


{\bibitem{gary1992} 
\Name{S P Gary} \REVIEW{\jgr}{97}{1992}{8523, doi: 10.1029/1992JA8523}.

\bibitem{gary1995} 
\Name{S P Gary, M F Thomsen, L Yin \and D Winske } \REVIEW{\jgr}{100}{1995}{21961, doi: 10.1029/1995JA21961}.}

\bibitem{treu2008} 
\Name{R A Treumann \and O D Constantinescu} \Year{2008} {arXiv: 0811.4328v1 [physics.space-ph].}

\bibitem{sahr2004}
\Name{F Sahraoui et al} \REVIEW{\ag}{22}{2004}{2283}.

\bibitem{hill1995}
\Name{P Hill et al}
\REVIEW{\jgr}{100}{1995}{9575, doi: 10.1029/94JA03194}.

\bibitem{kuz2007}
\Name{E A Kuznetsov,  T Passot \and P L Sulem}
\REVIEW{\prl}{98}{2007}{235003, doi: 10.1103/PhysRevLett.98.235003}.

{\bibitem{pokh2008} 
\Name{O A Pokhotelov et al} \REVIEW{\jgr}{113}{2008}{A04225, doi: 10.1029/2007JA012642}.}

\bibitem{calif2008}
\Name{F Califano et al}
\REVIEW{\jgr}{113}{2008}{Q08219, doi: 10.1029/2007JA012898}.

\bibitem{shoji2008}
\Name{M Shoji, Y Omura,  B T Tsurutani \and O P Verkho\-gly\-a\-dova} \Year{2008} 
{http://rp.iszf.irk.ru/hawk/URSI2008/ paper/HP04p3.pdf}.

{\bibitem{baum1999}
\Name{W Baumjohann et al}
\REVIEW{\ag}{17}{1999}{1528, doi: 10.1007/s005850050878}.}


\end{thebibliography}
\end{document}